\documentstyle[pre,preprint,aps]{revtex}
\tightenlines

\begin{document}
\draft

\title{
Dynamical mean-filed approximation to
small-world networks of spiking neurons: 
From local to global, and/or 
from regular to random couplings 
}
\author{
Hideo Hasegawa
\footnote{E-mail:  hasegawa@u-gakugei.ac.jp}
}
\address{
Department of Physics, Tokyo Gakugei University,
Koganei, Tokyo 184-8501, Japan
}
\date{\today}
\maketitle
\begin{abstract}
By extending a dynamical mean-field
approximation (DMA) previously proposed by the author
[H. Hasegawa, Phys. Rev. E {\bf 67}, 41903 (2003)],
we have developed a semianalytical theory 
which takes into account a wide range of couplings
in a small-world network.
Our network consists of
noisy $N$-unit FitzHugh-Nagumo (FN) neurons with  
couplings whose average coordination number $Z$ 
may change from local ($Z \ll N $) to global couplings ($Z=N-1$) 
and/or whose concentration of random couplings $p$ is allowed to vary
from regular ($p=0$) to completely random (p=1).
We have taken into account
three kinds of spatial correlations:
the on-site correlation, the correlation for a
coupled pair and that for a pair without direct couplings.
The original $2 N$-dimensional 
{\it stochastic} differential equations 
are transformed to 13-dimensional {\it deterministic}
differential equations expressed in terms
of means, variances and covariances of state variables. 
The synchronization ratio and the firing-time precision
for an applied single spike
have been discussed as functions of $Z$ and $p$.
Our calculations have shown that with increasing $p$,
the synchronization is {\it worse} because of increased
heterogeneous couplings, although the average
network distance becomes shorter. 
Results calculated by out theory are in good agreement with those 
by direct simulations.

\end{abstract}

\noindent
\vspace{0.5cm}
%
\section{INTRODUCTION}


It is well known that a brain forms complex networks
with nodes (neurons) and links (axons and dendrites).
A small patch of cortex may contain thousands of 
similar neurons, each connecting with hundreds
or thousands of other neurons in that same patch
or in other patches through axons and dendrites.
The underlying dynamics of individual neurons
is described by Hodgkin-Huxley-type nonlinear 
differential equations (DEs).  
Many theoretical studies have been reported on dynamics
of large-scale neuron networks.  
Extensive numerical calculations have been made by using
various spiking neuron models 
such as Hodgkin-Huxley (HH) \cite{Hodgkin52}, 
FitzHugh-Nagumo (FN) \cite{FitzHugh61,Nagumo62}
and Hindmarsh-Rose (HR) models \cite{Hindmarsh82}.   
These theoretical studies
have been performed with the use of the two approaches: direct simulations
and analytical methods such as the Fokker-Planck equation \cite{FPE},
the population density method \cite{PDM1,PDM2}
and the moment method \cite{Rod96,Tuckwell98,Rod98,Rod00}.
Since the computation time of direct simulations
is proportional to $N^2$, simulations for actual network size
become difficult, where $N$ is the size of a given neuron network. 
The Fokker-Plank equation method is mainly applied to $N=\infty$
network with the mean-field and/or diffusion approximations \cite{FPE2}.
The population method has been employed for a large-scale 
integrate-and-fire (IF) neuron network \cite{PDM1,PDM2}.
The moment method has been applied to FN and
HH neuron models \cite{Rod96,Tuckwell98,Rod98,Rod00}.

Most of theoretical studies have assumed that
couplings in neuron networks are local ($Z \ll N$)
or global ($Z=N-1$), and/or
regular ($p =0$) or random ($p=1$),
where $Z$ and $p$ denotes the average coordination number
and the concentration of random couplings, respectively.
In real neuron networks, however, 
couplings are neither local nor global
with the degree of
randomness locating between the two extremes of regular 
and random couplings.  
In recent years, much attention has been paid to 
{\it small-world} (SW) networks with the finite degree
of heterogeneity in couplings,
which is characterized by the high clustering and the small average
distance between nodes \cite{SW1,SW2,SW3,SW5}. 
The SW property is realized in
various kinds of biological, social and technological systems
such as the electric power grid, the movie-star collaborations
and the neuronal network of the nematode worm {\it C. elegans} 
\cite{SW1,SW2}.
Some calculations have been reported for
neural networks of spiking neuron models as well as of
phase models \cite{Fernandez00}-\cite{Hong02b}.
It has been shown
that by introducing the coupling heterogeneity into SW networks,
the synchronization is {\it increased} because the average
distance in SW networks is shorter than that in regular networks
\cite{Fernandez00,Barahona02,Bucolo02}
\cite{Kwon02,Hong02a,Hong02b}.
Recently, however, Nishikawa {\it et al.} \cite{Nishikawa03}  
have claimed that the synchronization is {\it decreased} with
including the coupling heterogeneity in SW networks.
Then it has been controversial whether the synchronization
in SW networks is better or worse than in regular networks.
These studies on SW networks have entirely relied on direct simulations,
and it is desirable to make a study by using an analytical method.

In previous papers of Refs. \cite{Hasegawa03a}
and \cite{Hasegawa03b} (which are referred to as I and II),
the present author proposed a semianalytical dynamical
mean-field approximation (DMA) theory for a study on
neuron ensembles (networks) with all-to-all (global) couplings.
In I, DMA was applied to an $N$-unit FN neuron network, 
for which $2N$-dimensional stochastic DEs 
are transformed to 8-dimensional deterministic DEs 
expressed by means, variances and covariances of state variables.
In the following II, DMA was applied to networks
consisting of general spiking neurons, each of which is
described by $M$ variables.
$MN$-dimensional stochastic DEs are transformed to
$N_{eq}$ deterministic DEs where $N_{eq}=M(M+2)$.
The DMA theory was successfully applied to
HH neuron network with $N_{eq}=24$ in II.
Advantages of DMA are 
(i) some qualitative properties of networks are derived
without 	numerical computations, and
(ii) the computational time of DMA is much
shorter than those of the moment method \cite{Note3}
and direct simulations. 
As for the item (ii), for example, the former is thousands times faster 
than the latter for $N=100$ HH neuron network
with 100 trials \cite{Hasegawa03b}. 

The purpose of the present paper is to 
develop a new approach for
SW neural networks of FN neurons with general couplings,
extending our semianalytical DMA \cite{Hasegawa03a}\cite{Hasegawa03b}.
In I and II, interactions among neurons
are assumed to be all-to-all (global) couplings.
For DMA to include local couplings in SW networks, 
we have taken into account variances and covariances which
express three kinds of spatial correlations:
(1) on-site correlation, (2) the correlation for a coupled pair and 
(3) that for an uncoupled pair without direct couplings. 
Assuming that the heterogeneity is small,
we have included its effects in order to
discuss the synchronization in SW networks.

The paper is organized as follows.
In Sec. II, we have derived DEs, applying the DMA
to SW networks consisting of FN neurons which are coupled
with the average coordination number $Z$. 
The original $2N$-dimensional stochastic DEs
are transformed to 13-dimensional deterministic DEs.
In Sec. IIIA, we report numerical calculations
for regular networks by changing $Z$ from local ($Z \ll N$) 
to global couplings ($Z=N-1$).
The $Z$-dependence of the firing-time accuracy and the synchronization
ratio for an applied single spike is discussed.
Numerical calculations for SW networks are reported in Sec. IIIB,
where the effect of the concentration of random couplings is discussed.
The final Sec. IV is devoted to conclusion and discussion. 

\section{Small-world networks of FN neurons}

\subsection{Adopted model and method}

We have assumed that $N$-unit FN neurons
are distributed on a ring 
with the average coordination number $Z$ and the concentration 
of random couplings $p$.
Dynamics of a single neuron $i$ in a given SW network
is described by the non-linear DEs given by 
\begin{eqnarray}
\frac{dx_{1i}(t)}{dt} &=& F[x_{1i}(t)]
- c \:x_{2i}(t)
+I_i^{(c)}(t)
+I_{i}^{(e)}(t)+\xi_i(t), \\
\frac{dx_{2i}(t)}{dt} &=& b \:x_{1i}(t) - d \:x_{2i}(t)+e,
\hspace{2cm}\mbox{($i=1$ to $N$)}
\end{eqnarray}
with
\begin{eqnarray}
I_i^{(c)}(t)&=& J \sum_j\: c_{ij}\:G(x_{1j}(t)), \\ 
I_i^{(e)}(t)&=& A \:\Theta(t-t_{in})\:\Theta(t_{in}+t_{w}-t).
\end{eqnarray}
In Eq. (1)-(4),
$F[x(t)]=k\: x(t)\: [x(t)-a]\: [1-x(t)]$, 
$k=0.5$, $a=0.1$, $b=0.015$, $d=0.003$ and $e=0$
\cite{Rod96}\cite{Tuckwell98}\cite{Hasegawa03a}: 
$x_{1i}$ and $x_{2i}$ denote the fast (voltage) variable
and slow (recovery) variable, respectively:
$G(x)$ stands for the sigmoid function given by
$G(x)=1/(1+exp[-(x-\theta)/\alpha])$
with threshold $\theta$ and width $\alpha$:
$J$ the coupling strength:
$c_{ij}$ the coupling factor given by $c_{ij}=c_{ji}=1$ for
a coupled $(i,\:j)$ pair and zero otherwise,
self-coupling terms being excluded ($c_{ii}=0$).
By changing $Z$ value, 
our model given by Eqs. (1)-(4) covers from local couplings ($Z \ll N$)
to global couplings ($Z=N-1$).
We have studied the response of neuron networks to
an external, single spike input given by $I_i^{(e)}(t)$
with magnitude $A$ and spike width $t_w$ applied at 
the input time $t_{in}$, $\Theta(x)$ being
the Heaviside function.
Added white noises $\xi_i(t)$ are given by
\begin{eqnarray}
<\xi_i(t)>&=&0, \\ 
<\xi_i(t)\:\xi_j(t')>&=&\beta^2 \; \delta_{ij}\:\delta(t-t'),
\end{eqnarray}
where the average of $<U({\bf z},t)>$
for an arbitrary function of $U({\bf z},t)$ is given by
\begin{equation}
<U({\bf z},t)>
= \int...\int \;d{\bf z} \;U({\bf z},t) \;Pr({\bf z}),
\end{equation}
$Pr({\bf z})$ denoting a probability distribution function 
for $2N$-dimensional
random variables ${\bf z}=(\{ x_{\kappa i} \})$.

An SW network is made after 
the Watts-Strogatz model \cite{SW1}.
Starting from the regular coupling 
for which $c_{ij} \equiv c_{0ij}$, 
$N_{ch}$ couplings among $N Z/2$ couplings
are randomly modified such that $c_{0ij}=0$ is changed to $c_{ij}=1$
or verse versa.
The concentration of random couplings is given by
\begin{equation}
p = \frac{2N_{ch}}{N Z},
\end{equation}
which is 0 and 1 for completely regular and random couplings,
respectively.
We shall take into account the effect of
the heterogeneity given by
\begin{equation}
\frac{\delta c_{ij}}{Z}=\frac{1}{Z}(c_{ij}-c_{0ij}),
\end{equation}
assuming it is small. 

After I, we will obtain equations of motions for means,
variances and covariances of state variables.
Variables spatially averaged over the ensemble are defined by
\begin{eqnarray}
X_{\kappa}(t)&=&\frac{1}{N}\;\sum_{i} \;x_{\kappa i},
\hspace{1cm}\mbox{$\kappa=1,\:2$}
\end{eqnarray}
and their means by
\begin{eqnarray}
\mu_{\kappa}(t)&=&\left< \left< 
X_{\kappa}(t) \right> \right>_c,
\end{eqnarray}
where the bracket $< \cdot >_c$ 
denotes the average over the coupling configuration.
As for variances and covariances of state variables,
we consider three kinds of spatial correlations:
(1) on-site correlation ($\gamma$),
(2) the correlation for a coupled pair ($\zeta$)
and (3) that for a pair without direct couplings ($\eta$):
\begin{equation}
\left< \left< \delta x_{\kappa i}
\: \delta x _{\lambda j}\right> \right>_c = \left\{
  \begin{array}{ll}
    \gamma_{\kappa,\lambda}, &\mbox{for $i=j$}\\
    \zeta_{\kappa,\lambda}, &\mbox{for $i\neq j,\; c_{ij}=1$}\\
    \eta_{\kappa,\lambda}, &\mbox{for $i \neq j,\; c_{ij}=0$}
  \end{array}\right.
\end{equation} 
where $\kappa, \lambda=1,\:2$ and
\begin{eqnarray}
\delta x_{\kappa i}(t)&=& x_{\kappa i}(t)-\mu_{\kappa}(t).
\end{eqnarray}
In Eq. (12), $\gamma_{\kappa,\lambda}$, 
$\zeta_{\kappa,\lambda}$ and 
$\eta_{\kappa,\lambda}$ are defined by
\begin{eqnarray}
\gamma_{\kappa,\lambda}(t)
&=&\left< \frac{1}{N} \sum_i \left<\delta x_{\kappa i}(t)
\:\delta x_{\lambda i}(t)\: \right> \right>_c, \\
\zeta_{\kappa,\lambda}(t)
&=& \left< \frac{1}{N Z} \sum_i \;\sum_{j}
c_{ij} \left< \delta x_{\kappa i}(t)
\;\delta x_{\lambda j}(t) \right> \right>_c, \\
\eta_{\kappa,\lambda}(t)
&=& \left< \frac{1}{N (N-Z-1)} \sum_i \;\sum_{j}
(1-\delta_{ij} - c_{ij}) 
\left< \delta x_{\kappa i}(t)
\;\delta x_{\lambda j}(t) \right> \right>_c.
\end{eqnarray}
For a later purpose, we define also the spatially-averaged
correlation given by
\begin{eqnarray}
\rho_{\kappa,\lambda}(t)&=& \left< \frac{1}{N^2} \sum_i \;\sum_{j}
\left< \delta x_{\kappa i}(t)
\;\delta x_{\lambda j}(t) \right> \right>_c,\\
&=& \left< \left<\delta X_{\kappa}(t) 
\:\delta X_{\lambda}(t) \right> \right>_c,
\end{eqnarray} 
where $\delta X_{\kappa}(t)=X_{\kappa}(t)-\mu_{\kappa}(t)$.
It is noted that $\gamma_{\kappa,\lambda}$, 
$\zeta_{\kappa,\lambda}$
$\eta_{\kappa,\lambda}$ and 
$\rho_{\kappa,\lambda}$ are not independent,
obeying the sum rule given by
\begin{equation}
N \rho_{\kappa,\lambda}
= \gamma_{\kappa,\lambda}+ Z \zeta_{\kappa,\lambda}
+ (N-Z-1) \eta_{\kappa,\lambda}.
\end{equation}
In order to derive Eqs. (14)-(19), 
we have employed the decomposition:
\begin{eqnarray} 
1&=&\delta_{ij}+(1-\delta_{ij})[c_{ij}+(1-c_{ij})], \nonumber \\
&=& \delta_{ij}+c_{ij}+(1-\delta_{ij}-c_{ij}), 
\end{eqnarray}
with $c_{ii}=0$.

In calculating means, variances and covariances given 
by Eqs. (11) and (14)-(17),
we have assumed that (1) the noise intensity is weak,
(2) the distribution of state variables takes the
Gaussian form, and (3) the coupling heterogeneity
of $\delta c_{ij}/Z$ is small. 
By using the first assumption, we expand DEs given by Eqs. (1)-(4)
in a power series of fluctuations around means.
The second assumption may be justified by some numerical calculations
for FN \cite{Tuckwell98}\cite{Tanabe01} and 
HH neuron models \cite{Tanabe99}\cite{Tanabe01a}. 
Based on the third assumption,
the effect of coupling fluctuations has been taken into account up to
the order of $O((\delta c_{ij}/Z)^2)$.

Before closing Sec. IIA, we briefly summarize the
introduced variables and their meanings as follows:
$N$, the number of neurons:
$Z$, the average coordination number:
$p$, the concentration of random couplings:
$J$, the coupling strength:
$c_{ij}$, the coupling factor between neurons $i$ and $j$:
$X_{\kappa}$, the spatially average of the fast ($\kappa=1$)
and slow ($\kappa=2$) variables;
$\mu_{\kappa}$, a mean value of $X_{\kappa}$;
$\gamma_{\kappa,\lambda}$,
$\zeta_{\kappa,\lambda}$, and
$\eta_{\kappa,\lambda}$, 
the correlations of on-site, a coupled pair
and an uncoupled pair, respectively.
Readers who are not interested in mathematical details, 
may skip to Sec. IIC where a summary of our method is presented.

\subsection{Equations of motions}

After some manipulations, we get the following DEs
(the  argument $t$ being suppressed;
for details, see appendix A):
\begin{eqnarray}
\frac{d \mu_1}{d t}&=&f_0 + f_2 \gamma_{1,1} -c \mu_2
+J Z (g_0 + g_1 \phi_1)
+I_{ext}, \\
\frac{d \mu_2}{d t}&=& b \mu_1 - d \mu_2 +e,  \\
\frac{d \gamma_{1,1}}{d t}&=& 2 (a \gamma_{1,1}- c \gamma_{1,2})
+ 2 J Z ( g_1 \zeta_{1,1}+  g_0 \phi_1)
+\beta^2, \\
\frac{d \gamma_{2,2}}{d t}&=& 2 (b \gamma_{1,2}- d \gamma_{2,2}),  \\
\frac{d \gamma_{1,2}}{d t}&=& b \gamma_{1,1}
+ (a-d) \gamma_{1,2} 
- c \gamma_{2,2}
+ J Z (g_1\zeta_{1,2}+ g_0 \phi_2),  \\
\frac{d \rho_{1,1}}{d t}&=& 2 (a \rho_{1,1} - c \rho_{1,2})
+  \left( \frac{2 J Z g_1}{N} \right) 
[\gamma_{1,1}+ZR \zeta_{1,1}+(N-ZR-1)\eta_{1,1}]
+\frac{\beta^2}{N}, \\
\frac{d \rho_{2,2}}{d t}&=& 2 (b \rho_{1,2}- d \rho_{2,2}),  \\
\frac{d \rho_{1,2}}{d t}&=& b \rho_{1,1}
+ (a-d) \rho_{1,2} - c \rho_{2,2}
+  \left( \frac{J Z g_1}{N} \right) 
[\gamma_{1,2}+ZR \zeta_{1,2}+(N-ZR-1)\eta_{1,2}], \\
\frac{d \zeta_{1,1}}{d t}&=& 2 (a \zeta_{1,1} - c \zeta_{1,2})
+  2 J g_1[\gamma_{1,1}
+ ZC \zeta_{1,1}+ (ZR-ZC-1) \eta_{1,1}],\\
\frac{d \zeta_{2,2}}{d t}&=& 2 (b \zeta_{1,2}- d \zeta_{2,2}),  \\
\frac{d \zeta_{1,2}}{d t}&=& b \zeta_{1,1}
+ (a-d) \zeta_{1,2} - c \zeta_{2,2}
+  J g_1 [\gamma_{1,2} 
+ ZC \zeta_{1,2}+(ZR-ZC-1) \eta_{1,2}],\\
\frac{d \eta_{1,1}}{d t}&=& 2 (a \eta_{1,1} - c \eta_{1,2})\nonumber \\
&+&  \left ( \frac{2 J Z g_1}{N-Z-1} \right) 
\left\{ g_1[(ZR-ZC-1)\zeta_{1,1}+ (N-2ZR+ZC) \eta_{1,1}] 
- g_0 \phi_1 \right\}, \\
\frac{d \eta_{2,2}}{d t}&=& 2 (b \eta_{1,2}- d \eta_{2,2}),  \\
\frac{d \eta_{1,2}}{d t}&=& b \eta_{1,1}
+ (a-d) \eta_{1,2} - c \eta_{2,2} \nonumber \\
&&+  \left( \frac{J Z }{N-Z-1} \right) 
\left\{ g_1 [(ZR-ZC-1)\zeta_{1,2}
+ (N-2ZR+ZC) \eta_{1,2}] - g_0 \phi_2 \right\}, \\
\frac{d \phi_1}{d t}&=& a \phi_1- c \phi_2
+ J Z g_0 \:\delta R_p, \\
\frac{d \phi_2}{d t}&=& b \phi_1- d \phi_2,
\end{eqnarray}
with 
\begin{eqnarray}
\phi_{\kappa}(t) &=& \left< \frac{1}{NZ} \sum_i \sum_j
\left< \delta x_i \delta c_{ij} \right> \right>_c,
\hspace{1cm}\mbox{$\kappa=1,\;2$} \\
C&=& \frac{1}{NZ^2} \sum_i \sum_j \sum_k \:c_{0ij}c_{0jk}c_{0ik},\\
R&=& \frac{1}{N Z^2} \sum_i \sum_j \sum_k c_{0ij}\:c_{0jk}, \\
\delta R_p&=& \left< \frac{1}{NZ^2} \sum_i \sum_j \sum_k 
\:\delta c_{ij} \delta c_{jk} \right>_c,
\end{eqnarray}
where $a=f_1+ 3 f_3 \gamma_{1,1}$,
$f_{\ell}= (1/\ell !) F^{(\ell)}$,
$g_{\ell}= (1/\ell !) G^{(\ell)}$,
$C$ corresponds to the clustering coefficient introduced in
SW networks \cite{SW1,SW2}, $R$ expresses 
the coupling connectivity, and
$\delta R_p$ is its fluctuation part,
related discussions being given in Sec. IV.

\subsection{Summary of our method}

Equations of motions for 
$\mu_{\kappa}(t)$, $\gamma_{\kappa,\lambda}(t)$, 
$\zeta_{\kappa,\lambda}(t)$, $\eta_{\kappa,\lambda}(t)$
and $\rho_{\kappa,\lambda}(t)$
are given by Eqs. (21)-(40). In Eqs. (35) and (36), 
$\phi_{\kappa}(t)$ ($\kappa=1,\:2$) are new correlation functions
which appear in the process of calculating equations of motion
of $\gamma_{\kappa,\lambda}$ {\it et al.}
The factors $C$, $R$ and $\delta R_p$
defined by Eqs. (38)-(40) generally depend on the geometry of
a given neuron network.
For a regular ring with even $Z$, we get 
$R=1$ and $C$ given by 
\begin{equation}
C= \left\{
 \begin{array}{ll}
   0, &\quad\mbox{for $Z \leq 2$}\\
   3/4-3/2Z, &\quad\mbox{for $4 \leq Z <  2N/3$} \\
   3/4-3/2Z +9/4-(3N-9/2)/Z \\
      \;\;\; +(N^2-3N+2)/Z^2, &\quad\mbox{for $2N/3 \leq Z < N-1$} \\
   (1-1/Z).&\quad\mbox{for $Z = N-1$}
 \end{array}\right.
\end{equation}
Figure 1 shows $C$ as a function of $Z/N$ 
for $N=100$, 200, 500 and 1000.
We note
that $C \sim 0.75$ for $0.1 < Z/N < 0.7$ and that
$C \rightarrow (1-1/Z)$ as $Z/N \; \rightarrow \; (1-1/N)$.
In the case of global couplings ($Z =N-1$),
however, we get $C = (1-1/Z)$ independent of the geometry.
$\delta R_p$ defined by Eq. (40), which 
expresses fluctuations in heterogeneous couplings, is
increased with increasing
the concentration of random couplings, $p$ [Fig. 6(a)].
Among the 12 correlations such as $\gamma_{\kappa,\lambda}$ {\it et al.}
given by Eqs. (14)-(17), 
9 correlations are independent because of the
sum rule given by Eq. (20). In this study,
we have chosen nine correlations of $\gamma_{\kappa,\lambda}$,
$\zeta_{\kappa,\lambda}$ and $\rho_{\kappa,\lambda}$
as independent variables.
Then the original $2 N$-dimensional {\it stochastic}
DEs given by Eqs.(1) and (2) have been transformed to
13-dimensional {\it deterministic} DEs.

It is worthwhile to explain the relation
between the present theory and I
where the original $2N$-dimensional stochastic DEs
for regular, global couplings 
are transformed to 8-dimensional deterministic DEs.
In the present study for the general coupling, 
we have to take into account
$\zeta_{\kappa,\lambda}$ and $\eta_{\kappa,\lambda}$, in order to 
discriminate correlations between a coupled pair
and an uncoupled pair.
However, in the limit of
$Z=N-1$ for regular, global couplings for which $R=1$ and  $ZC=Z-1$,
$\eta_{\kappa,\lambda}$ are not necessary because 
there are no uncoupled pairs:
prefactors of $(ZR-ZC-1)$ for $\eta_{\kappa,\lambda}$ 
in Eqs. (32) and (34) vanish with
$\phi_{\kappa}=0$.
Then the number of required DEs is reduced from 13 to 8.
Equations (21)-(28) for $\mu_{\kappa}$, $\gamma_{\kappa,\lambda}$
and $\rho_{\kappa,\lambda}$
agree with Eqs. (20)-(27) in I \cite{Note1}.

\subsection{Firing-time accuracy and synchronization}


\noindent
{\bf Firing-time accuracy}

When we solve DEs given by Eqs. (21)-(36), 
we may obtain various quantities
relevant to firings in neuron networks.
The firing time of a given neuron $i$ is defined 
as the time when the variable
$x_{1i}(t)$ crosses the threshold $\theta$ from below:
\begin{equation}
t_{o\ell} = \{t \mid x_{1i}(t)=\theta; \dot{x}_i(t) > 0 \}.
\end{equation}
It has been shown that the distribution of firing times of $t_{o\ell}$
is given by \cite{Hasegawa03a}
\begin{eqnarray}
Z_{\ell}(t) &\sim& 
\Phi \left(\frac{t-t_f}{\delta t_{o\ell}} \right)
\frac{d}{dt} \left(\frac{\mu_1}{\sqrt{\gamma_{1,1}(t_f)}} \right) 
\Theta(\dot{\mu}_1),\\
&\rightarrow& \delta{(t-t_f)}, 
\hspace{2cm}\mbox{for $\gamma_{1,1}(t_f) \:\rightarrow \:0$} \nonumber
\end{eqnarray}
with
\begin{eqnarray}
\delta t_{o\ell} &=& \frac{\sqrt{\gamma_{1,1}(t_f)}}{\dot{\mu}_1}, 
\end{eqnarray}
where $\Phi$ expresses the normal distribution function,
the average firing time
$t_f$ is implicitly defined by $\mu_1(t_f) = \theta$,
$\dot{\mu}_1=\dot{\mu}_1(t_f)$
and the dot denotes the time derivative. 

Similarly, the firing time of an averaged variable $X_1(t)$ is defined 
as the time when the variable
$X_1(t)$ crosses the threshold $\theta$ from below:
\begin{equation}
t_{og} = \{t \mid X_1(t)=\theta; \dot{X}_1(t) > 0 \}.
\end{equation}
The distribution of firing times of $t_{og}$
is given by \cite{Hasegawa03a}
\begin{eqnarray}
Z_{g}(t) &\sim& 
\Phi \left(\frac{t-t_f}{\delta t_{og}} \right)
\frac{d}{dt} \left(\frac{\mu_1}{\sqrt{\rho_{1,1}(t_f)}} \right) 
\Theta(\dot{\mu}_1), \\
&\rightarrow& \delta{(t-t_f)}.
\hspace{2cm}\mbox{for $\rho_{1,1}(t_f) \:\rightarrow \:0$} \nonumber
\end{eqnarray}
with
\begin{eqnarray}
\delta t_{og} &=& \frac{\sqrt{\rho_{1,1}(t_f)}}{\dot{\mu}_1}. 
\end{eqnarray}

\vspace{0.5cm}

\noindent
{\bf Synchronization ratio}

We discuss the synchronization in neuron networks,
considering the quantity given by
\begin{eqnarray}
R_s(t)&=& \frac{1}{N^2} 
\sum_i \sum_{j} <[x_i(t)-x_j(t)]^2>,\\
&=& 2 (\gamma_{1,1}-\rho_{1,1}),
\end{eqnarray}
which vanishes in the completely synchronous state.
From a comparison of Eqs. (23)-(25) with Eqs. (26)-(28),
we note that
\begin{eqnarray}
\rho_{\kappa,\lambda} &=& \frac{\gamma_{\kappa,\lambda}}{N},
\hspace{1cm}\mbox{for $J \rightarrow 0$} 
\end{eqnarray}
Then, $R_s(t)$ given by Eq. (49) becomes
$R_s(t)=(1-1/N)\gamma_{1,1}(t) \equiv R_{s0}(t)$
in the asynchronous state,
while $R_s(t)=0$ in the completely synchronous state.
We define the {\it synchronization ratio} at
the firing time $t_f$ by \cite{Hasegawa03a}
\begin{equation}
S_f=S(t_f),
\end{equation}
with
\begin{equation}
S(t)=1 - \frac{R_s(t)}{R_{s0}(t)}
= \left( \frac{N\rho_{1,1}(t)/\gamma_{1,1}(t)-1}{N-1} \right),
\end{equation}
which is 0 and 1 for completely asynchronous and
synchronous states, respectively.
The synchronization ratio shows much variety depending on model parameters such
as the coupling strength ($J$), the noise intensity ($\beta$), 
the size of cluster ($N$), the coordination number ($Z$),
and the random concentration ($p$), as will be discussed in Sec. III.

\section{CALCULATED RESULTS}

\subsection{Regular couplings}

We have adopted same parameters of $\theta=0.5$,
$\alpha=0.5$, $\tau_s=10$, $A=0.10$, $t_{in}=100$
and $T_w=10$ as in I \cite{Hasegawa03a}.
DMA calculations have been made by solving Eqs. (21)-(36) with
the use of the fourth-order Runge-Kutta method with
the time step of 0.01.
We have performed direct simulations by using also  
the fourth-order Runge-Kutta method with the time step of 0.01.
Results of direct simulations are averages of 1000 trials for
$Z \le 20$ (or $N \le 20$) and those of 100 trials otherwise noticed.
All quantities are dimensionless.

First we discuss the case of regular couplings ($p=0$),
by changing the average coordination number $Z$
from local ($Z \ll N$) to global couplings ($Z=N-1$).
The plots in Figs. 2(a)-2(c) show firings in an $N=100$ neuron 
network with regular couplings 
for $Z=10$, 50 and 99 with $\beta=0.01$ and $J=0.002$ 
when a single external spike given by Eq. (4) is applied.
Figures 2(a)-2(c) show that as increasing $Z$, 
scattering of firing times is reduced, which suggests
an increase in the firing accuracy and the synchronization.
These are results of direct simulations with single trials.
They are more clearly discussed 
with calculations using the DMA.
Fig. 2(d)-2(f) show time courses of
$S(t)$ calculated in the DMA for $Z=10$, 50 and 99, whose magnitudes
are increased as increasing $Z$;
note differences of the ordinate scales in Figs. 2(d)-2(f).
The synchronization ratio at firing times, $S_f$,
is 0.0019, 0.0113 and 0.0295 for $Z=10$, 50 and 99, respectively,
which shows an increased synchrony with increasing $Z$.

We will discuss some details of the DMA calculation in 
Figs. 3(a)-3(d) which show time courses of $\mu_1$, $\gamma_{1,1}$,
$\zeta_{1,1}$, and $\rho_{1,1}$, respectively,
for regular couplings ($p=0$) with $\beta=0.01$, $J=0.002$,
$N=100$ and $Z=10$.
Results of DMA expressed by solid curves are in
good agreement with those of direct simulations
depicted by dashed curves.
Time courses of $\mu_1$, 
$\gamma_{1,1}$ and $\rho_{1,1}$ shown in Fig. 3(a), 3(b) and 3(d)
for local couplings (Z=10) are not so different from those for
global couplings having been reported in Fig. 1 of I,
except for their magnitudes.
For example, DMA calculations for the local coupling with $Z=10$
in the case of $\beta=0.01$, $J=0.002$ and $N=100$ show that
magnitudes of $\gamma_{1,1}$, $\zeta_{1,1}$ and $\rho_{1,1}$
at the firing time of $t=104.44$ are
0.271 $\times 10^{-2}$, 0.475 $\times 10^{-4}$ 
and 0.320 $\times 10^{-4}$, respectively.
In contrast, for the global coupling with $Z=99$,
magnitudes of $\gamma_{1,1}$, $\zeta_{1,1}$ and $\rho_{1,1}$
at the firing time of $t=103.88$ are
0.235 $\times 10^{-2}$, 0.693 $\times 10^{-4}$
and $0.921 \; \times \;10^{-4}$, respectively.

Figure 4(a) shows the $Z$ dependence of
$\gamma_{1,1}$, $\zeta_{1,1}$ and $\rho_{1,1}$
at the firing time
with $J=0.002$, $\beta=0.01$ and $N=100$;
filled and open marks express results of DMA and direct simulations,
respectively.
Results of $\gamma_{1,1}$ and $\rho_{1,1}$ of DMA are indistinguishable
from those of direct simulations.
With increasing $Z$, 
both $\zeta_{1,1}$ and $\rho_{1,1}$ are increased, 
while $\gamma_{1,1}$ is slightly decreased,
as mentioned above.
The $Z$ dependence of the firing time $t_f$ 
is plotted in Fig. 4(b), which
shows the faster response for larger $Z$.
This is due to the fact that by an increased $Z$,
$\mu_1$ is increased more rapidly to cross the threshold level of $\theta$.
Then $\dot{\mu}_1$ at $t=t_f$ is increased with increasing $Z$, as
the chain curve in Fig. 4(c) shows.
Figure 4(c) shows that with increasing $Z$,
the firing-time accuracy of $\delta t_{o\ell}$
is improved while that of $\delta t_{og}$ is independent of $Z$.
The $Z$ dependence of the synchronization is plotted in Fig. 4(d)
showing $S_f$ to be linearly increased for a small $Z$.
This clearly explains the larger synchrony $S_f$ for larger $Z$, 
having been shown in Figs. 2(a)-2(f).

\subsection{SW couplings}

Next we discuss the case of SW couplings, by changing the
concentration of random couplings $p$.
The plots in Figs. 5(a)-5(c) show firings in SW 
networks for $p=0.0$, 0.1 and 1.0, respectively, with $\beta=0.005$
$J=0.02$, $N=100$ and $Z=10$ calculated by
direct simulations with single trials,
when a single external spike given by Eq. (4) is applied.
In this subsection,
we have adopted a smaller $\beta$ and a larger $J$ than in Sec. IIIA
to get more evident effects of $p$.
Figures 5(a)-5(c) show that as increasing $p$, 
scattering of firing times is gradually increased, which suggests
a decrease in the firing accuracy and the synchronization.
These results are more clearly seen in calculations with the use of DMA.
Fig. 5(d)-5(f) show time courses of
$S(t)$ for $p=0$, 0.1 and 1.0, calculated in the DMA.
The synchronization ratio at firing times $S_f$
is 0.0256, 0.0224, and 0.0114, for $p=0$, 0.1 and 1.0, respectively.
Although $S_f$ for $p=0.1$ is nearly equal to that for $p=0.0$,
the time course of $S(t)$ for $p=0.1$ is rather different from
that for $p=0.0$. 

This decrease in $S_f$ with increasing $p$ mainly arises from 
an increased $\delta R_p$, as shown in
Fig. 6(a) where the $p$ dependence of $\delta R_p$ is plotted
for $Z=10$, 20 and 50 of a given ring with $N=100$.
With increasing $p$, $\delta R_p$ is linearly increased 
as $\delta R \propto p/Z$ for a small $p$.
Figure 6(b) will be explained in Sec. IV.

Figure 7(a) shows the $p$ dependence of
$\gamma_{1,1}$, $\zeta_{1,1}$ and $\rho_{1,1}$
at the firing time
with $J=0.02$, $\beta=0.005$, $N=100$ and $Z=10$;
filled and open marks express results of DMA and direct simulations,
respectively.
At $p=0.0$, $\gamma_{1,1}$, $\zeta_{1,1}$ and $\rho_{1,1}$
are $0.671 \;\times 10^{-3}$, $0.131 \;\times 10^{-3}$
and $0.239 \;\times 10^{-4}$, respectively.
In contrast, at $p=1.0$, they
are $0.109 \;\times 10^{-2}$, $0.144 \;\times 10^{-3}$
and $0.232 \;\times 10^{-4}$, respectively.
With increasing $p$, 
$\gamma_{1,1}$ is increased,
while $\rho_{1,1}$ and $\zeta_{1,1}$ are almost constant.
The difference between the $p$ dependences of $\gamma_{1,1}$,
$\rho_{1,1}$ and $\zeta_{1,1}$ arises from the fact that
$d \gamma_{1,1}/dt$ in Eq. (23) has a contribution from
$\phi_1$ while $d \rho_{1,1}/dt$ and $d \zeta_{1,1}/d t$ 
in Eqs. (26) and (29) have no direct
contributions from it.
Figure 7(b) shows that the firing time of $t_f=103.88$ 
is independent of $p$, which is in accordance with
a constant $\dot{\mu}_1$ shown in Fig. 7(c).
Figure 7(c) shows that with increasing $p$,
the firing-time accuracy of $\delta t_{o\ell}$
becomes worse because of an increased $\gamma_{1,1}$
while that of $\delta t_{og}$ is independent of $p$.
The $p$ dependence of $S_f$ is depicted in Fig. 7(d),
which shows that the synchrony is decreased with increasing $p$.
This clearly explains results of smaller $S_f$
for larger $p$, having been shown in Figs. 5(a)-5(f). 

\section{CONCLUSION AND DISCUSSION}


Generalizing a phenomenological analysis
adopted in I \cite{Hasegawa03a}
based on calculated results of DMA, we have tried to get
an analytical expression for $S_f$. 
From calculated results discussed in the previous section, 
we expand $\gamma_{1,1}$ and $\rho_{1,1}$ in a series of
$JZ$ and $p$:
\begin{eqnarray}
\gamma_{1,1} &=& \gamma_0 [1-a_1 JZ(1-a_2 p) +\cdot \cdot  ], \\
\rho_{1,1} &=& \frac{\gamma_0}{N} (1+b_1 J Z +\cdot \cdot), 
\end{eqnarray}
where $\gamma_o \propto \beta^2$, 
and $a_1$, $a_2$ and $b_1$ are positive coefficients.
We have obtained an expression for $\gamma_{1,1}$ given by Eqs. (53),
because the effect of $p$ should vanish for $J=0$ or $Z=0$.
Substituting Eqs. (53) and (54) to Eq. (52), we get
\begin{eqnarray}
S_f &=& \left( \frac{a_1(1-a_2p)+b_1)}{N-1} \right) J Z 
+ \cdot \cdot.
\end{eqnarray}
The expression for $S_f$ given by Eq.(55) well explains the behavior
shown in Figs. 4(d) and 7(d). 
Dependences of the quantities on $N$, $Z$, $J$ and $\beta$
for local couplings are the same as those for all-to-all
couplings having discussed in I.
Typical examples of $N$ dependence of various quantities
are shown in Figs. 9(a)-9(d).
Figures 9(a) and 9(b) show that $\rho_{1,1} \propto N^{-1}$ while 
$\gamma_{1,1}$, $\zeta_{1,1}$ and $t_f$ are independent of $N$,
which yields $\delta t_{o\ell} \propto N^{-1/2}$ and 
$\delta t_{og} \propto N^0$, as shown in Fig. 9(c).
Figure 9(d) shows that $S_f \propto N^{-1}$ both for local and global
couplings, expressing that the synchronization is more easily realized
in smaller networks than in larger ones.  

In an early stage of this study, we obtained DEs
given by Eqs. (21)-(34) with $\phi_1=\phi_2=0$, but
with $C$ and $R$ which are replaced by $C_p$ and $R_p$, respectively, 
given by
[for detail see after Eq. (A22) in appendix A]
\begin{eqnarray}
C_p&=& \left< \frac{1}{NZ^2} \sum_i \sum_j \sum_k 
\:c_{ij}c_{jk}c_{ik} \right>_c, \\
R_p&=& \left<  \frac{1}{N Z^2} \sum_i \sum_j \sum_k 
c_{ij}\:c_{jk} \right>_c.
\end{eqnarray}
In this formulation, the effect of the couplings heterogeneity
is included in the $p$-dependent clustering coefficient $C_p$ and 
coupling connectivity $R_p$.
The clustering coefficient $C_p$ denotes an averaged fraction
for given three nodes to be mutually coupled \cite{SW1,SW2}. 
The $p$ dependence of $C_p$ is depicted in Fig, 6(b) which shows 
that with increasing $p$,
$C_p$ is decreased and approaches $C_p = Z/N$ at $p = 1$.
In contrast, the coupling connectivity $R_p$ expresses an averaged fraction
for given two nodes, which are not necessarily coupled,
to have a common neighboring node.   
$R_p$ in Eq. (57) may be rewritten as 
\begin{eqnarray}
R_p
&=&\frac{1}{Z^2}\;\sum_K K^2 P(K)  
\equiv \frac{1}{Z^2}\:\overline{K^2},
\end{eqnarray} 
where the overline denotes the average over
$P(K)$ expressing the probability for a given neuron to have
$K$ couplings \cite{Note2}. It is easy to see that $R_p$ 
is given by $R_p=1 + \delta R_p$ [Eqs. (60) and (61)], the $p$ dependence of
$\delta R_p$ being plotted in Fig. 6(a).
Unfortunately results calculated with the use of
$C_p$ and $R_p$ for finite $p$ were not in
good agreement with those of direct simulations
because effects of coupling heterogeneity are not properly taken
into account in such DEs.

After several tries, we have obtained DEs having been given 
by Eqs. (21)-(36). $C$, $R$ and $\delta C_p$ given by Eqs. (38)-(40)
may be expressed in terms of $C_p$ and $R_p$ as \cite{Note2}
\begin{eqnarray}
C&=&C_0, \\
R&=& R_0 =1, \\
\delta R_p&=& R_p -R_0=\frac{1}{Z^2}\overline{(K-\overline{K})^2},
\end{eqnarray}
with $Z=\overline{K}$.
Figure 8 shows $P(K)$
for $p=0.0$, 0.1, 0.2 and 1.0 with $N=100$ and $Z=10$.
In the limit of $p=0$, $P(K)\;(=\delta_{K,Z})$ is 
the delta function.
With increasing $p$, $P(K)$ has the distribution
centered at $K=Z$.
In the limit of $p=1$,
$P(K)$ approaches the Poisson distribution \cite{SW5}.
Figure 6(a) shows that with increasing $p$, 
$\delta R_p$ is increased, while $C_p$ is decreased 
as shown in Fig. 6(b).
An increased $\delta R_p$
yields an increase in $\gamma_{1,1}$, by which $S_f$ is decreased
and $\delta t_{o\ell}$ is increased.
It should be noted that 
effects of heterogeneous couplings are taken into account
by $\delta R_p$ through the correlation
functions $\phi_1$ and $\phi_2$ in Eqs. (35) and (36),
which play important roles in dynamics of SW networks.

To summarize, 
we have developed a semianalytical theory
for SW networks of spiking FN neurons,
including three kinds spatial correlations:
correlations of on-site, a coupled pair
and an uncoupled pair.
By changing $Z$ and $p$, we have performed model calculations
of the response of the network to an external single spike.
It has been shown that 

\noindent
(1) when $Z$ is increased, 
the synchronization ratio $S_f$ and
the firing-time accuracy $\delta t_{o\ell}$ are improved 
[Figs. 4(c) and 4(d)], which arises from
a decrease in $\gamma_{1,1}$ and an increase in
$\rho_{1,1}$, and

\noindent
(2) when $p$ is increased, 
both $S_f$ and $\delta t_{o\ell}$ become worse [Figs. 7(c) and 7(d)]
due to an increase in $\gamma_{1,1}$
induced by fluctuations in the coupling heterogeneity.

\noindent
The item (1) is easily understood.
The result for $S_f$ in the item (2) is consistent 
with that of Ref. \cite{Nishikawa03}.
It, however, contradicts some calculations 
\cite{Fernandez00,Barahona02,Bucolo02}\cite{Kwon02,Hong02a,Hong02b},
which have claimed that the synchronization in SW networks
is better than that in regular networks,
since communication between neurons
is more efficient because of the shorter characteristic path length $L$
(as for the $p$-dependence of $L$, see Fig. 2 of Ref. \cite{SW1}).
Our semianalytical theory with the use of the DMA,
which is valid for weak noise ($\beta \ll 1$) 
and small coupling heterogeneity ($\delta R_p \ll 1$),
has shown that the synchrony of SW networks depends on
$R$, $C$ and $\delta R_p$ given by Eqs. (38)-(40), 
but it is not affected by the average path length, $L$.
In particular, $\delta R_p$, $\phi_1$ and $\phi_2$ have been shown to play
crucial roles in dynamics of SW neural networks. 
Although the item (2) discussed above 
relies on the definition of the synchronization
ratio of $S(t)$ given by Eq. (52), this conclusion is not changed
even if we adopt an alternative measure for the synchrony.
For example, when we employ $R_s$ given by Eq. (49), $R_s$ is increased
with increasing $p$ because of an increased $\gamma_{1,1}$,
which again signifies the worse synchronization in SW networks.
The semianalytical theory developed in this paper
can be applied not only to SW neural networks 
but also to a wide class of complex SW networks.
When we apply our theory to a general SW network
in which dynamics of each node is 
described by $M$-dimensional stochastic DEs, 
we get $N_{eq}$-dimensional deterministic DEs
where $N_{eq}=M(3M+7)/2$. For example, $N_{eq}=5$ for
Langevin model ($M=1$), $N_{eq}=13$ for FN model ($M=2$),
and $N_{eq}=38$ for HH model ($M=4$).
The items (1) and (2) [and also Eq. (55)] which have been
derived for FN neuron model,
are expected to hold for any SW network.

The present approach shares in its advantages with the original DMA
previously proposed in I: 
(i) some results may be derived without numerical calculations
because of its semianalytical nature, and
(ii) a computational time for a large-scale system by DMA is 
much shorter than that by direct simulations.
By extending the ring geometry adopted in this paper,
we may discuss the response of more realistic
synfire-chain-type SW networks
\cite{Hasegawa03a}\cite{Abeles93}. 
In the present paper, we have neglected the transmission time delay.
Because the average path length $L$
becomes shorter by the appearance of shortcuts
\cite{SW1,SW2,SW3,SW5},
the response speed is expected to be improved 
in SW networks with time delays.
Recently, we have successfully
applied the DMA to stochastic ensembles
with time-delayed regular couplings
\cite{Hasegawa04a,Hasegawa04b}. By using our approach,
we may discuss dynamics of general SW networks 
with time delays within the framework of the DMA.
In the so-called {\it scale-free} (SF) network
such as the world-wide web and the network of
citations of scientific papers,
the link connectivity $P(K)$ for a node to interact 
to $K$ other nodes follows
a power-law distribution $P(K) \sim K^{-\gamma}$
with the index $\gamma$ ($\sim 2.1$ to $4$) \cite{Bara99},
in contrast to an exponential 
distribution for a large $K$ in our SW networks (Fig. 8).
This SF distribution probability originates from the two factors,
the growth of nodes and their preferential
attachment \cite{Bara99}. 
Quite recently it has been reported that the functional connectivity
$P(K)$ versus the distance $K$ in human brain 
is given by a SF distribution: 
$P(K) \sim K^{-2}$ \cite{Chialvo04}.
It is interesting to apply our semianalytical 
approach to such SF networks.
These subjects raised above are left as our future study.

\section*{Acknowledgements}
This work is partly supported by
a Grant-in-Aid for Scientific Research from the Japanese 
Ministry of Education, Culture, Sports, Science and Technology.  


\appendix

\section{Derivation of Eqs. (21)-(36)}


Substituting Eqs. (9) and (13)
to Eqs. (1)-(4),
we get DEs for $\delta x_{1i}$ and $\delta x_{2i}$ of a neuron $i$,
given by (argument $t$ is suppressed)
\begin{eqnarray}
\frac{d \delta x_{1i}}{d t}
&=& f_1 \delta x_{1i}+f_2 (\delta x_{1i}^2-\gamma_{1,1})
+ f_3 \delta x_{1i}^3 - c \delta x_{2i}  
+ \delta I_i^{(c)} +\xi_j, \\
\frac{d \delta x_{2j}}{d t}&=& b \delta x_{1j} - d \delta x_{2j},
\end{eqnarray}
with
\begin{eqnarray}
\delta I_i^{(c)}(t)
&=& J \:\sum_{j}  [g_1(t) c_{0ij}\delta x_{1j}(t)
+ g_0(t) \delta c_{ij} + g_1(t) \delta c_{ij} \delta x_{1j}(t)+ \cdot], 
\end{eqnarray}
where $f_{\ell}= (1/\ell !) F^{(\ell)}$ and $g_{\ell}= (1/\ell !) G^{(\ell)}$. 
DEs for the correlations are given by
\begin{eqnarray}
\frac{d \gamma_{\kappa,\lambda}}{d t}
&=& \left< \frac{1}{N} \sum_{i}
\left< \left[ \delta x_{\kappa i}\:
\left(\frac{d \delta x_{\lambda i}}{d t}\right)
+\left( \frac{d \delta x_{\kappa i}}{d t} \right)
\:\delta x_{\lambda i} \right] \right> 
\right>_c, \\
\frac{d \zeta_{\kappa,\lambda}}{d t}
&=& \left< \frac{1}{N Z} \sum_{i} \sum_{j} c_{ij} 
\left< \left[ \delta x_{\kappa i} \:
\left( \frac{d \delta x_{\lambda j}}{d t} \right)
+\left( \frac{d \delta x_{\kappa j}}{d t} \right)
\:\delta x_{\lambda i} \right] \right>
\right>_c,\\
\frac{d \rho_{\kappa,\lambda}}{d t}
&=& \left< \frac{1}{N^2} \sum_{i} \sum_{j} 
\left< \left[ \delta x_{\kappa i} \:
\left( \frac{d \delta x_{\lambda j}}{d t} \right)
+\left( \frac{d \delta x_{\kappa j}}{d t} \right) 
\:\delta x_{\lambda i} \right] \right>
\right>_c.
\end{eqnarray}
With the use of Eqs. (A1)-(A3), 
we may calculate DEs given by Eqs. (21)-(34).
For example, terms including $\delta I_i^{(e)}$ 
in $d \gamma_{1,1}/d t$,
$d \zeta_{1,1}/d t$ and $d \rho_{1,1}/d t$ become
\begin{eqnarray}
\left< \frac{2 }{N} \sum_i \left< \delta x_{1i} 
\delta I_i^{(c)} \right> \right>_c 
&=& \frac{2 J}{N} \sum_i \sum_j 
g_1 c_{0ij} \left< \left<\delta x_{1i} \delta x_{1j} 
\right> \right>_c \nonumber\\
&+& \frac{2 J}{N} \sum_i \sum_j  
g_0 \left< \left< \delta x_{1i} \delta c_{ij} \right> \right>_c, \\
&=& 2 J Z [g_1 \zeta_{1,1}+ g_0 \phi_1],\\
\left< \frac{2}{NZ} \sum_i \sum_j
c_{ij} \left< \delta x_{1i} \delta I_j^{(c)} \right> \right>_c 
&=& \frac{2 J}{NZ} 
\sum_i \sum_j \sum_k
g_1 c_{0ij} c_{0jk} \left< \left< \delta x_{1i} \delta x_{1k} \right> 
\right>_c \nonumber \\
&+& \frac{2 J}{NZ} 
\sum_i \sum_j \sum_k
g_0 c_{0ij} \left< \left< \delta x_{1i} \delta c_{jk} \right> \right>_c,\\
&=& 2 J g_1[\gamma_{1,1}+ZC \zeta_{1,1}+(ZR-ZC-1)\eta_{1,1}], \\
\left< \frac{2}{N^2} \sum_i \sum_j 
\left< \delta x_{1i} \delta I_j^{(c)} \right> \right>_c 
&=&  \frac{2 J}{N^2} \sum_i \sum_j \sum_k
g_1 c_{0jk} \left< \left< \delta x_{1i} \delta x_{1k} \right> 
\right>_c \nonumber \\
&+&  \frac{2 J}{N^2} \sum_i \sum_j \sum_k
g_0 \left< \left< \delta x_{1i} \delta c_{jk} \right> \right>_c,\\
&=& \frac{2 J Z g_1}{N}[\gamma_{1,1}+Z R\zeta_{1,1}+(N-ZR-1) \eta_{1,1}],
\end{eqnarray}
where $\phi_{\kappa}$ ($\kappa=1,2$)
are new correlation functions defined by
\begin{eqnarray}
\phi_{\kappa}(t) &=& \left< \frac{1}{NZ} \sum_i \sum_j
\left< \delta x_{\kappa i}(t) \delta c_{ij} \right> \right>_c.
\hspace{1cm}\mbox{$\kappa=1,\;2$}
\end{eqnarray}
In evaluating Eqs. (A7)-(A12), 
we have employed the relations given by
\begin{eqnarray}
1&=& \frac{1}{NZ}\sum_i \sum_j c_{0ij},\\
R&=& \frac{1}{NZ^2}\sum_i \sum_j \sum_k c_{0ij} \:c_{0jk},\\
C&=&\frac{1}{NZ^2}\sum_i \sum_j \sum_k c_{0ij} \:c_{0jk} \:c_{0ik},\\
\delta R_p &=& \frac{1}{NZ^2}
\left< \sum_i \sum_j \sum_k \delta c_{ij} \:\delta c_{jk} \right>_c,
\end{eqnarray}
and the mean-field approximation given by
\begin{eqnarray}
\left< \left< \delta x_{\kappa i} \delta \:x_{\lambda j} \right> \right>_c
&=& \gamma_{\kappa,\lambda}\:\delta_{ij}
+ (1-\delta_{ij})\:[\zeta_{\kappa,\lambda}\:\delta_{ij} \:c_{ij} 
+ \eta_{\kappa,\lambda}\:\delta_{ij} \:(1-c_{ij})], \\
&=& \gamma_{\kappa,\lambda}\:\delta_{ij} 
+  \zeta_{\kappa,\lambda}\:c_{ij}\:  
+ \:\eta_{\kappa,\lambda}\:(1-\delta_{ij}-c_{ij}) ,\\
\left< \left< \delta x_{\kappa,i} \delta c_{jk} \right> \right>_c
&=&  \phi_{\kappa} c_{jk}\:(\delta_{ij}+\delta_{ik}),
\end{eqnarray}
with the Gaussian decoupling
approximations \cite{Hasegawa03a}.
In Eqs. (A18) and (A19), $\gamma_{\kappa,\lambda}$, $\zeta_{\kappa,\lambda}$
and $\eta_{\kappa,\lambda}$ denote the correlations of on-site,
a coupled pair and an uncoupled pair, which are defined by Eqs. (12)-(14). 
The approximations given by Eqs. (A18)-(A20) are consistent with
the definition of $\gamma_{\kappa,\lambda}$, 
$\zeta_{\kappa,\lambda}$ 
and $\eta_{\kappa,\lambda}$ given by Eqs. (14)-(16),
and those of $\phi_{\kappa}$ given by Eq. (37).

The equations of motion of $\phi_{\kappa}$ are similarly
calculated with the use of the relation given by
\begin{eqnarray}
\frac{d \phi_{\kappa}}{d t}
&=& \left< \frac{1}{NZ} \sum_i \sum_j
\left< \left( \frac{d \delta x_{\kappa i}}{d t}\right) 
\:\delta c_{ij} \right> \right>_c,
\end{eqnarray}
which yield Eqs. (35) and (36).

We have taken into account terms up to orders of
$O((\delta x)^2)$, $O((\delta c/Z)^2)$ and $O(\delta x \:\delta c/Z)$
in Eqs. (21)-(36),
and up to the order of $O((\delta x)^4)$ in the term including 
$a\:(=f_1+3 f_3 \gamma_{1,1})$
which plays an important role in stabilizing DEs \cite{Hasegawa03a}.

On the contrary, when we adopt an expression given by
\begin{eqnarray}
\delta I_i^{(c)}(t)
&=& J \:\sum_{j}  [g_1(t) c_{ij}\delta x_{1j}(t)+ \cdot], 
\end{eqnarray}
instead of Eq. (A3), DEs given 
by Eqs. (A7), (A9) and (A11) become
\begin{eqnarray}
\left< \frac{2 }{N} \sum_i \left< \delta x_{1i} 
\delta I_i^{(c)} \right> \right>_c 
&\simeq&  \frac{2 J}{N} \sum_i \sum_j 
g_1 \left< c_{ij} \left<
\left<\delta x_{1i} \delta x_{1j} \right> \right>_c \right>_c, \\
&=& 2 J Z g_1 \zeta_{1,1},\\
\left< \frac{2}{NZ} \sum_i \sum_j
c_{ij} \left< \delta x_{1i} \delta I_j^{(c)} \right> \right>_c 
&\simeq&  \frac{2 J}{NZ} 
\sum_i \sum_j \sum_k
g_1 \left< c_{ij} c_{jk} 
\left< \left< \delta x_{1i} \delta x_{1k} \right> \right>_c
\right>_c,\\
&=& 2 J g_1[\gamma_{1,1}+ZC_p \zeta_{1,1}+(ZR_p-ZC_p-1)\eta_{1,1}],\\
\left< \frac{2}{N^2} \sum_i \sum_j 
\left< \delta x_{1i} \delta I_j^{(c)} \right> \right>_c 
&\simeq&  \frac{2 J}{N^2} \sum_i \sum_j \sum_k
g_1 \left< c_{jk}  
\left< \left< \delta x_{1i} \delta x_{1k} \right> \right>_c
\right>_c,\\
&=& \frac{2 J Z g_1}{N}[\gamma_{1,1}
+Z R_p\zeta_{1,1}+(N-ZR_p-1) \eta_{1,1}],
\end{eqnarray}
where decoupling approximations such as
\begin{equation}
\left< c_{ij} \left< 
\delta x_{1i}\:\delta x_{1j} \right> \right>_c
\simeq
\left< c_{ij} \left< \left< 
\delta x_{1i}\:\delta x_{1j} \right>\right>_c \right>_c,
\end{equation}
and Eq. (A19) are employed.
$C_p$ and $R_p$ in Eqs. (A26) and (A28) 
are given by Eqs. (56) and (57).
Note that $c_{ij}$ in Eqs. (A23), (A25) and (A27)
depends on the configuration of couplings while
$c_{0ij}$ in Eqs. (A7), (A9) and (A11) does not.
Then we got equations of motions given by
Eqs. (21)-(34) with $\phi_1=\phi_2=0$ but with
$C$ and $R$ which are, respectively, 
replaced by $p$ dependent $C_p$ and $R_p$
given by Eqs. (56) and (57).
As mentioned in Sec. IV, results calculated 
with the use of such DEs are not in good agreement
with those obtained by direct simulations
because effects of coupling fluctuations are not properly included
in the formulation mentioned above.
It is indispensable to take into account effects 
of the coupling heterogeneity expressed by $\delta R_p$ 
through the correlation functions 
$\phi_1$ and $\phi_2$, as given by Eqs. (35) and (36).



\begin{figure}
\caption{
The clustering coefficient $C$ for 
a ring with regular couplings ($p=0$)
as a function of $Z/N$ 
for $N=100$, 200, 500 and 1000. 
}
\label{fig1}
\end{figure}

\begin{figure}
\caption{
(color online).
The plots showing firings in a regular neuron network
for (a) $Z=10$, (b) 50 and (c) 99
calculated by direct simulations (single trials), and
time courses of $S(t)$
for (d) $Z=10$, (e) 50 and (f) 99
calculated by DMA (solid curves) 
and direct simulations (dashed curves)
($\beta=0.01$, $J=0.002$, $N=100$ and $p=0.0$).
Arrows in (d)-(f) denote firing times.
}
\label{fig2}
\end{figure}

\begin{figure}
\caption{
(color online).
Time courses of (a) $\mu_1$, (b) $\gamma_{1,1}$, (c) $\zeta_{1,1}$,
and (d) $\rho_{1,1}$
for $\beta=0.01$, $J=0.002$, $N=100$, $Z=10$ and $p=0$,
solid and dashed curves denoting results of DMA
and direct simulations, respectively.
At the bottom of (a), an input signal is plotted. 
}
\label{fig3}
\end{figure}

\begin{figure}
\caption{
(color online).
The $Z$ dependence of 
(a) the correlations of $\gamma_{1,1}$ (circles), 
$\zeta_{1,1}$ (triangles) 
and $\rho_{1,1}$ (squares) at the firing time,
(b) the firing times $t_{f}$,
(c) the firing-time accuracy of $\delta t_{o\ell}$ (circles)
and $\delta t_{og}$ (squares) and $\dot{\mu}_1$ (triangles), and
(d) the synchronization ratio at the firing time, $S_f$,
for $\beta=0.01$, $J=0.002$ and $N=100$:
filled and open marks denote results of DMA and
direct simulations, respectively.
Results of $\zeta_{1,1}$ and $\dot{\mu}_1$ are only for DMA.
}
\label{fig4}
\end{figure}

\begin{figure}
\caption{
(color online).
The plots showing firings in a small-world neuron network
for (a) $p=0.0$, (b) 0.1 and (c) 1.0
calculated by direct simulations (single trials), and
time courses of $S(t)$
for (d) $p=0.0$, (e) 0.1 and (f) 1.0
calculated by DMA (solid curves) 
and direct simulations (dashed curves)
($\beta=0.005$, $J=0.02$ and $N=100$).
Arrows in (d)-(f) denote firing times.
}
\label{fig5}
\end{figure}

\begin{figure}
\caption{
The $p$ dependence of (a) the factor $\delta R_p$ and
(b) the clustering coefficient $C_p$,
for $Z=10$, 20 and 50 with $N=100$.
}
\label{fig6}
\end{figure}

\begin{figure}
\caption{
(color online).
The $p$ dependence of 
(a) the correlations of $\gamma_{1,1}$ (circles), 
$\zeta_{1,1}$ (triangles)
and $\rho_{1,1}$ (squares) at the firing time,
(b) the firing times $t_{f}$,
(c) the firing-time accuracy of $\delta t_{o\ell}$ (circles)
and $\delta t_{og}$ (squares), and $\dot{\mu}_1$ (triangles), and
(d) the synchronization ratio at the firing time, $S_f$,
for $\beta=0.005$, $J=0.02$, $N=100$ and $Z=10$:
filled and open marks denote results of DMA and
direct simulations, respectively. 
Results of $\zeta_{1,1}$ and $\dot{\mu}_1$ are only for DMA.
}
\label{fig7}
\end{figure}

\begin{figure}
\caption{
The probability $P(K)$ for a given neuron to have
$K$ couplings for $p=0.0$, 0.1, 0.2 and 1.0 
with $N=100$ and $Z=10$ in a SW ring .
}
\label{fig8}
\end{figure}

\begin{figure}
\caption{
(color online).
The $N$ dependence of 
(a) the correlations of $\gamma_{1,1}$ (circles), 
$\zeta_{1,1}$ (triangles)
and $\rho_{1,1}$ (squares) at the firing time,
(b) the firing times $t_{f}$,
(c) the firing-time accuracy of $\delta t_{o\ell}$ (circles)
and $\delta t_{og}$ (squares), and
(d) the synchronization ratio at the firing time, $S_f$
($\beta=0.01$, $J=0.002$, $N=100$ and $Z=10$):
filled and open marks denote results of DMA and
direct simulations, respectively.
In (d) results for global couplings ($Z=N-1$) are also shown.
}
\label{fig9}
\end{figure}

\end{document}